\documentclass{emulateapj}
\usepackage[below]{placeins}
\usepackage{graphics,graphicx} 
\usepackage{natbib}

\slugcomment{Submitted to Astrophysical Journal}

\def\gtorder{\mathrel{\raise.3ex\hbox{$>$}\mkern-14mu
    \lower0.6ex\hbox{$\sim$}}}
\def\ltorder{\mathrel{\raise.3ex\hbox{$<$}\mkern-14mu
    \lower0.6ex\hbox{$\sim$}}}

\def\msun{M_\odot}

\newcommand {\rs} {$R_{\rm s}$}
\newcommand {\rt} {$R_{\rm t}$}

\newcommand {\mt} {$M_{\rm t}$}

\newcommand {\rvir} {$R_{\rm vir}$}
\newcommand {\mvir} {$M_{\rm vir}$}

\newcommand {\rvmax} {$R_{\rm vmax}$}

\shorttitle{Substructure Evolution in Dark Matter and Baryonic Halo Models}
\shortauthors{Romano-Diaz et al.}

\begin{document}

\title{Dissecting Galaxy Formation:\\
II. Comparing Substructure in Pure Dark Matter and Baryonic Models}

\author{ 
Emilio Romano-D\'{\i}az\altaffilmark{1},
Isaac Shlosman\altaffilmark{1},
Clayton Heller\altaffilmark{2},
Yehuda Hoffman\altaffilmark{3}
}
\altaffiltext{1}{
Department of Physics and Astronomy, 
University of Kentucky, 
Lexington, KY 40506-0055, 
USA
}
\altaffiltext{2}{
Department of Physics, 
Georgia Southern University, 
Statesboro, GA 30460, 
USA
}
\altaffiltext{3}{
Racah Institute of Physics, Hebrew University; Jerusalem 91904, Israel
}

\begin{abstract}
We compare the substructure evolution in pure dark matter (DM) halos with
those in the presence of baryons, hereafter PDM and BDM models. The prime
halos have been analyzed in the previous work, Romano-Diaz et al. Models
have been evolved from identical initial conditions which have been
constructed by means of the Constrained Realization method. The BDM model
includes star formation and feedback from stellar evolution onto the gas.
A comprehensive catalog of subhalo populations has been compiled and
individual and statistical properties of subhalos analyzed, including their
orbital differences. We find that
subhalo population mass functions in PDM and BDM are consistent with a 
single power law, $M_{\rm sbh}^{\alpha}$, for each of the models in the 
mass range of $\sim 2\times 10^8~{\rm M_\odot} - 
2\times 10^{11}~{\rm M_\odot}$. However, we detect a nonnegligible
shift between these functions, the time-averaged $\alpha\sim -0.86$
for the PDM and $-0.98$ for the BDM models. Overall, $\alpha$ appears
to be a nearly constant with variations of $\pm 15\%$. Second,
we find that the radial mass distribution of subhalo populations can
be approximated by a power law, $R^{\gamma_{\rm sbh}}$ with a steepening that 
occurs at the radius of a maximal circular velocity, \rvmax, in the prime 
halos. Here we find that the $\gamma_{\rm sbh}\sim -1.5$ for the PDM and
--1 for the BDM models, when averaged over time inside \rvmax. The slope 
is steeper outside this region and approaches $-3$. We detect little 
spatial bias (less than $10\%$) between the subhalo populations and the 
DM distribution of the main halos. Also, the subhalo population exhibits 
much less triaxiality in the presence of baryons, in tandem with the
shape of the prime halo. Finally, we 
find that, counter-intuitively, the BDM population is depleted at a faster 
rate than the PDM one within the central 30~kpc of the prime halo. The 
reason for this is that although the baryons provide a substantial glue 
to the subhalos, the main halo exhibits the same trend. This assures a more
efficient tidal disruption of the BDM subhalo population. However, this 
effect can be reversed for a more efficient feedback from stellar 
evolution and the central supermassive black holes, which will expel
baryons from the center and decrease the central concentration of the
prime halo. We compare our results with via Lactea and Aquarius 
simulations and other published results.
\end{abstract}

\keywords{cosmology: dark matter --- galaxies: evolution --- galaxies:
formation --- galaxies: halos --- galaxies: interactions --- galaxies:
kinematics and dynamics}
    
\section{Introduction}
\label{sec:intro}

The high-redshift Universe is characterized by a uniform mixture of dark
matter (DM) and baryons (e.g., Spergel et al. 2007). The process of
galaxy formation is expected to lead essentially to a (partial)
separation between the DM and baryons, due to dissipative processes in
the latter. As a result, the central regions of many galaxies should
be dominated by the baryons and by their dynamics --- this is
supported by observations of various aspects of galactic dynamics
(e.g., Flores \& Primack 1994; de Blok \& Bosma 2002; 
Sand et al. 2004; Gentile, Tonini \& Salucci 2007; de Blok et al. 2008). 
On the other hand and as a direct consequence of the process
of a hierarchical buildup of structure in the Universe, a large amount
of substructure, i.e., subhalos, penetrating the prime halos is
expected to be present (e.g., Klypin et al. 1999;
Moore et al. 1999).  Moreover, the baryons are expected to modify the
DM structure on sub-galactic scales, both in the prime halos and the
subhalos, although the extent of this change is far from being clear. 
Such differences can be accompanied by substantial adjustments
in the DM distribution, angular momentum, and other dynamic
variables. The baryonic processes may also affect
the survival of subhalos. This in turn can modify the basic parameters
of the galactic disks which grow within DM halos, because disk-subhalo
interactions can drive the disk evolution (e.g., Gauthier, Dubinski \& Wilson
2006; Heller, Shlosman \& Athanassoula 2007a,b; Romano-Diaz et al. 2008b; 
Shlosman 2010 and refs. therein). These effects are far from being fully 
quantified --- a direct
comparison between pure DM and DM$+$baryon models on subgalactic
scales is limited both by numerical resolution and complex baryonic
physics.  In Paper~I (Romano-Diaz et al. 2009), we compared the
evolution of the prime halos of pure DM and DM$+$baryon numerical
(hereafter PDM and BDM) models, evolved from identical initial
conditions within the cosmological framework.  Here we compare some
aspects of the evolution in the substructure associated with the prime
halos in the PDM and BDM models.

Recent efforts to understand galaxy formation have been
spearheaded by pure DM halo formation in the cosmological framework,
culminating in the large-scope {\it Millenium} (Springel et al. 2005), 
{\it via Lactea} (Diemand, Kuhlen \& Madau 2007; Diemand et al. 2008),
{\it Aquarius} (Springel et al.  2008) and {\it Ghalo} (Stadel et al. 
2009) simulations. Addition of a baryon component to
the pure DM modeling is associated with difficulties related to a
numerical modeling of dissipative processes, partly due to the
sub-grid physics, and to our limited knowledge of physics of the ISM,
star formation, energy and momentum feedback, etc., as shown in
DM$+$hydrodynamic simulations  (e.g., Sommer-Larsen, G\"otz \& Portinari 
2003; Maccio et al. 2006; Berentzen \& Shlosman 2006;
Stinson et al. 2006; Governato et al. 
2007; Heller et al. 2007a; Kaufmann et al. 2007;
Romano-Diaz et al. 2009; Scannapieco et al. 2009),
in chemodynamical simulations (e.g., Samland \& Gerhard 2003; Brook et al. 
2007a,b), or implementing semianalytical methods (e.g., Scannapieco et
al. 2006). Comparison between
various aspects of subhalo population between pure DM and baryonic
models has been performed also for cluster scales (e.g., Nagai \& 
Kravtsov 2005; Weinberg et al. 2008; Dolag et al. 2009).

In Paper~I and Romano-Diaz et al. (2008a,b) we quantified to what
extent the baryons alter the DM density distribution within the prime
halo. They contribute decisively to the evolution of its central
region, leading initially to an isothermal DM cusp, which is
subsequently flattened to a DM density core --- the result of heating
by dynamical friction of the DM$+$baryon subhalos during the quiescent
evolution epoch. This confirmed previous work on this subject (e.g.,
El-Zant, Shlosman \& Hoffman 2001; El-Zant et al. 2004; Tonini, Lapi
\& Salucci 2006; see also review by Primack 2009, as well as Johansson, 
Naab \& Ostriker 2009).  As a by-product of
this process, the cold gas has been ablated from a growing embedded
disk, reducing the star formation rate by a factor of 10, and heating
up the spheroidal gas and stellar components, triggering their
expansion. We find that only a relatively small $\sim 20\%$ fraction
of DM particles in PDM and BDM models are bound within the radius of
maximal circular velocity in the halo, most of the DM particles
perform larger radial excursions.  We also find that the fraction of
baryons within the halo virial radius somewhat increases during the
major mergers and decreases during the minor mergers. The net effect
appears to be negligible --- an apparent result of our choice of
feedback from stellar evolution. Furthermore, we find that the DM
halos are only partially relaxed beyond their virialization. While the
substructure is being tidally disrupted, mixing of its debris in the
halo is not efficient and becomes even less so with redshift. The
phase-space correlations (streamers) formed after $z\sim 1$ survive
largely to the present time.

This paper is structured as follows. Section~2 describes the initial
conditions and the numerical modeling. Section~3 focuses on the basic
properties of the PDM and BDM subhalos in our simulations, and
section~4 compares the evolution of these subhalo populations in PDM 
and BDM prime halos. Discussion and conclusions are give in the last 
section.

\section{Initial Conditions and Numerical Modeling}

\begin{figure*}
\begin{center}
\includegraphics[angle=0,scale=0.61]{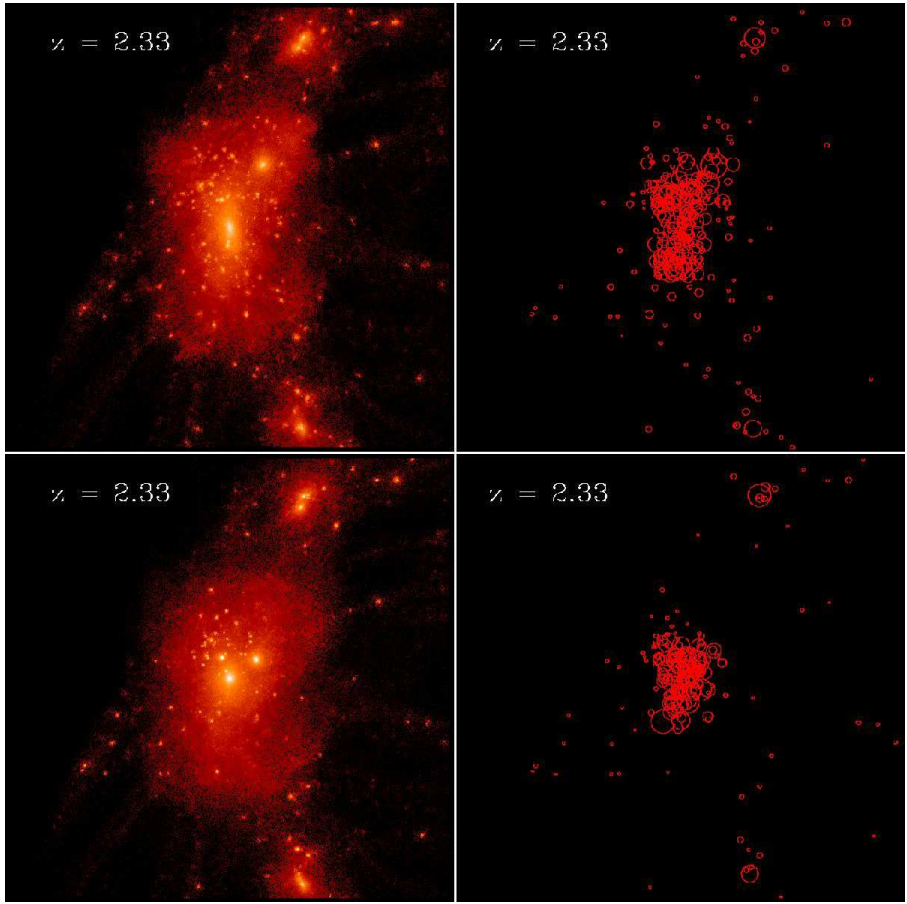}
\end{center}
\caption{Snapshots of PDM prime halo (top left) with the associated 
subhalo population (top right) and the DM in BDM (bottom left)
prime halos with the associated subhalos (bottom right) at $z=2.33$. 
The circle sizes scale with the tidal masses of the subhalos. The box 
size is 600~kpc. The colors correspond to the local DM mass 
density. \mvir{} and \rvir{} of the main halos are 
$\sim 2\times 10^{12}~{\rm M_\odot}$ and $\sim 160$~kpc, for PDM and
BDM models. The tidal masses of three most massive subhalos are 
$(8.8, 4.3, 3.7)\times 10^{10}~{\rm M_\odot}$ and 
$(9.3, 3.3, 2.4)\times 10^{10}~{\rm M_\odot}$, for PDM and BDM 
respectively.
}
\end{figure*}

For numerical details, model parameters and initial conditions, the
reader is referred to Paper~I. Here we only comment that the initial
conditions for PDM and BDM models have been obtained using Constrained
Realizations method (CRs, Bertschinger 1987; Hoffman \& Ribak 1991;
van de Weygaert \& Bertschinger 1996). We followed the prescription of
Hoffman \& Ribak (1991) to build the initial conditions within a
restricted cubic volume of space with sides $L = 8h^{-1}$Mpc in the
$\Lambda$CDM cosmology, where a sphere of $5h^{-1}$~Mpc is carved out
and evolved from $z = 120$. Numerical simulations have been performed 
with the parallel version of FTM-4.5 hybrid $N$-body/Smooth Particle 
Hydrodynamics (SPH) code (e.g., Heller \& Shlosman 1994; Heller et al. 
2007a; Paper~I) using vacuum boundary conditions and physical
coordinates. The star formation (SF) modeling algorithm is described
in Heller et al. (2007a) and involves multiple generations of stars,
energy and momentum feedback from stellar evolution, supernovae and OB
stellar winds.

The total mass inside the computational sphere is $\sim 6.1\times
10^{12}~h^{-1}M_\sun$. To introduce the baryons, we have randomly replaced
DM particles on the initial conditions grid by baryon (i.e., SPH)
particles, so that $\Omega_{\rm m}$ stays the same.  The initial
masses of DM and SPH particles are the same, $2.78\times
10^6~M_\sun$. The evolution of various parameters characterizing the
DM and baryons has been followed in 1000 snapshots, linearly spaced in
the cosmological expansion parameter $a$.

\subsection{Identifying and Measuring the Subhalos}

The HOP algorithm (Eisenstein \& Hut 1998) used to identify the
substructure within the computational sphere becomes inefficient
deep within the prime halo, where density rises above the virial
density $\rho_{\rm vir}\equiv \Delta(z) \rho(z)$. Here $\Delta(z)$ is
the critical overdensity at virialization (Bryan \& Norman 1998) and 
$\rho(z)$ is the
background density. The halo virial radius, \rvir, is defined
in the context of the spherical top-hat collapse model,
\mvir $= 4/3\pi \Delta(z) \rho(z)$\rvir$^3$.
Therefore, we divide the computational sphere into
two regions --- the boundary between these regions is \rvir{} of a
prime halo.  In the outer, mostly unvirialized part we simply apply
the HOP algorithm for halo finding. In the inner, mostly virialized
part, we refine the method and perform the following iterative
procedure.

\begin{itemize}

\item We start from \rvir{} and split the enclosed volume into
arbitrary-shaped shells lying between isodensity surfaces, where
$\rho_{\rm i+1}= \eta \rho_{\rm i}$, with the shells counted from
\rvir{} inwards.  We choose $\eta = 1.5$. Within each shell, the total
(DM or DM$+$baryons) background density was taken as constant and
equal to the average of two isodensities which sandwich it.

\item The HOP algorithm was applied within each such shell, the
subhalos have been identified and their tidal radii, \rt{}, and
masses, \mt{} have been determined from the condition that the subhalo
density falls below the background (shell) density, $\rho_{\rm i}$.
We check that such substructures are repeated in several
iterations. Otherwise they are considered spurious and removed from
the subhalo catalog.

\item In order to distinguish ``real" structures from (unbound) density
enhancements, we require that real substructure will have a peak
density $\rho > \eta \rho_{\rm i}$ {\it and} that the number of DM
particles associated with this candidate exceeds 100. We find that
objects that include less than the above number of DM particles
provide unreliable statistical data in order to analyze their
properties.

\item Finally, the background (i.e., unbound to a subhalo) DM or
baryon particles have been assessed using the velocity histogram of
all particles within \rt{} in order to separate them from the genuine
subhalo mass contribution. We use velocities in the center of mass 
(CoM) of a
subhalo. An alternative method exists to separate the bound subhalo
gas from the hot unbound background, using the bi-modal temperature
distribution of the gas particles.

\end{itemize} 
 
As stated above,
we also define the virial and tidal radii of subhalos. For this we use
the 3D density contours of subhalos obtained in the previously
described procedure. The last contour still belonging to the subhalo
is that whose density is defined in the first item. It is 
above the local background density of the
prime halo (if located inside the prime halo) or simply the background
density (if outside the prime halo).
The tidal radius of the subhalo,
\rt, is defined as the maximal extent of such a contour from the
subhalo CoM. The virial radius of a subhalo is defined similarly when
the latter is situated outside the prime halo.  The subhalo masses,
both virial and tidal, are assumed to lie within density shells of these 
radii,
respectively.  Snapshots of prime halos and the associated subhalos
are shown in Fig.~1 for PDM and BDM models at $z=2.33$.

\section{Results: Basic Properties of Subhalos in PDM and BDM Models}

We have compiled a comprehensive catalog of subhalos in PDM and BDM
models which is used to infer their basic properties. In the
following, we also choose representative samples of PDM and BDM
subhalos out of the general catalog having different masses (i.e., in
the mass ranges of $10^8\msun-10^9\msun, 10^9\msun-10^{10}\msun,
10^{10}\msun-10^{11}\msun$) at the time of crossing \rvir{} of prime
halos.  Finally, we look at the corresponding subhalo pairs in PDM and
BDM.

Our tests indicate that only subhalos with particle number exceeding
100 give a reliable estimate of their properties (see also Trenti et al.
2010).
In order to identify the time of a subhalo (tidal) dissolution, we
mark the 20 densest DM particles in each of the subhalos and calculate the
dispersion in their relative positions, $\sigma(z)$, as a function of
time (section~4). The tidal disruption of a subhalo is reflected in a
sharp increase in $\sigma$. We choose $\sigma=5$~kpc for the time when
the subhalo has been destroyed, and mark this time as $z_{\rm des}$ or
$a_{\rm des}$.  This method has been tested and found as very
reliable.

Because we avoid fitting the NFW density profile to the subhalos, we
modified the definition for the compactness parameter, $c$, used in
the literature by replacing \rs{} by \rvmax{} and \rvir{} by \rt. This
has been done by obtaining the rotation curves for the subhalos based
on their total mass distribution. In Paper~I we have shown that
\rvmax{} behaves largely similar to \rs. The new definition is
$c=$\rt/\rvmax.  Because \rvmax{} is the most bound radius (i.e.,
maximum of the DM rotation curve), we typically find that $c\gtorder
1$ over the lifetime of a subhalo, i.e., it is destroyed before $c$
drops below unity.
  
\subsection{Subhalo Mass Function}
 
\begin{figure}
\begin{center}
\includegraphics[angle=0,scale=0.4]{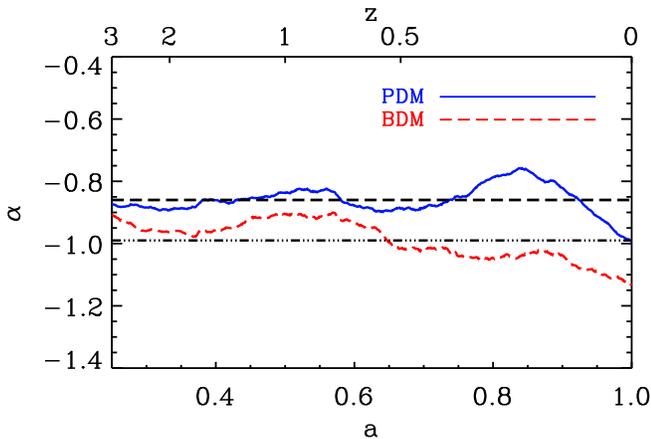}
\end{center}
\caption{Evolution of the SubHMF, $N_{\rm sbh}(M_{\rm sbh})$, in PDM
(blue, solid) and BDM (red, dashed) models using a power-law fit,
$N(M_{\rm sbh})\sim M_{\rm sbh}^{\alpha}$. Power index $\alpha$
within \rvir{} is shown as a function of redshift $z$ and the
cosmological expansion factor $a$. Only subhalos with $N_{\rm
sbh}\ge 100$ have been used. This corresponds to $M_{\rm sbh}\gtorder 
2.7\times
10^8~M_\sun$. The values of $\alpha$ at $z=0$ are --0.99 and --1.13
for PDM and BDM respectively. We have smoothed the curves averaging
them over the time frames. The averages of both curves are shown as
a black dashed line at -0.86 (PDM) and a black dot-dashed one at 
-0.98 (BDM).  }
\end{figure}

\begin{figure}
\begin{center}
\includegraphics[angle=0,scale=0.5]{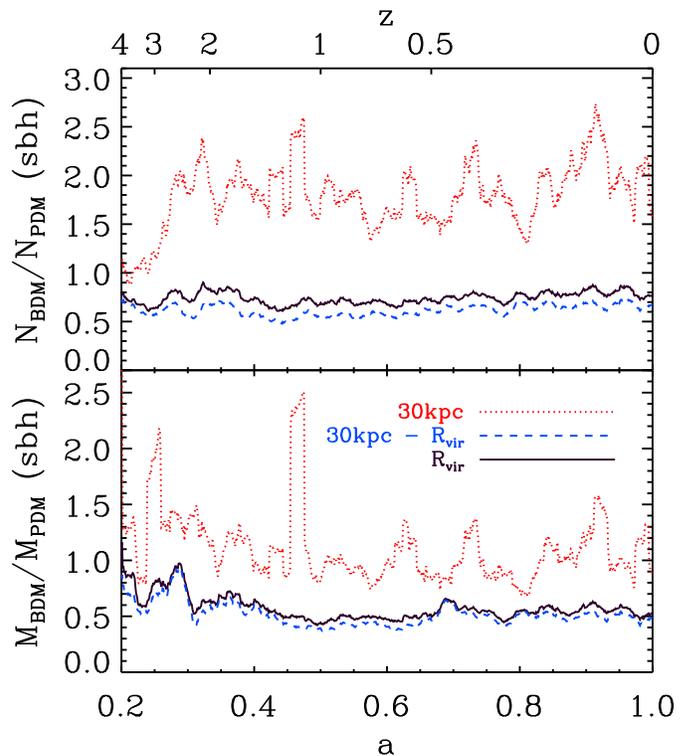}
\end{center}
\caption{Evolution of the SubHMF in PDM and BDM models: ratio of
subhalo numbers (top) and masses (bottom) inside 30~kpc, within 30~kpc
--- \rvir, and inside \rvir{} of the prime halos.  Only subhalos with
$N_{\rm DM} \ge 100$ have been used. All curves have been
time-averaged over 50 frames. }
\end{figure}

We follow the evolution of the subhalo mass function, $N(M_{\rm sbh})$
inside \rvir{} of the prime halos, hereafter SubHMF.  We approximate
the SubHMF with a power law, $N_{\rm sbh}\sim M_{\rm sbh}^{\alpha}$
(e.g., Moore et al. 1999; Ghigna et al. 2000; Springel et al. 2001),.
We start by comparing the evolution of the power law index $\alpha$ in
the SubHMF (Fig.~2).

The main conclusions that emerge from Fig.~2 with regards to the
evolution of $\alpha$ in PDM and BDM models are as follows. First,
the PDM SubHMF remains slightly shallower than the BDM one at all
times, which is reflected in their averages as well, --0.86 vs --0.98
respectively. The final values at $z=0$ are -0.99 and -1.13. Second,
the difference between these models, $\Delta \alpha(z)$, otherwise
nearly constant, appears to increase and decrease after $z\sim
0.5$. During this time, the PDM SubHMF becomes slightly shallower,
while the BDM one slightly steeper. Both functions converge
thereafter. Third, the associated fluctuations in $\alpha(z)$ with
respect to the mean values of $\alpha$ are related to the waves of
inflowing low mass subhalos across \rvir{} after the epoch of major
mergers (Romano-Diaz et al. 2008a). Overall, this behavior is
consistent with $\alpha\sim $~const. over the whole period, in the
presence of an intrinsic scatter of $\pm 15\%$.

In principle, we cannot rule out that there is an evolutionary trend in
$\alpha$ for $z\ltorder 0.2$  in both models in Fig.~2. It may be 
also related to the overall depletion of the subhalo population, i.e.,
the small number statistics, which would allow larger fluctuations
from the mean toward the end of the simulations. As noted in the Discussion
section, a similar trend was detected in the analysis of pure DM 
simulations of via Lactea (Madau et al. 2008). 

To compare the evolution of the subhalo populations within \rvir{} and
in the innermost 30~kpc in PDM and BDM, we display their number and
mass ratios as a function of time in Fig.~3. Within the central
30~kpc, the BDM-to-PDM {\it number} ratio of subhalos stays above 
unity for
the entire evolution after $z\sim 3$, while prior to $z\sim 3$ it
oscillates in a narrow range around unity.  After $z\sim 1.5$, this
ratio is limited to $1.5-2.5$ range and shows large amplitude
fluctuations due to the clumpy influx (see above and in Romano-Diaz et
al. 2008a), especially noticeable in the coresponding mass ratios in
the lower frame of Fig.~3. On the other hand, this ratio remains
steady and below unity in the less dense outer parts of the prime
halos. This is also true for the subhalo population overall.

The DM {\it mass} ratio, BDM/PDM, of subhalos (Fig.~3) within the inner 
30~kpc appears to stabilize around unity at later times, albeit 
fluctuating with substantial amplitudes.  Beyond this region, i.e., 
between 30~kpc -- \rvir{}, it stabilizes around 0.5 -- 0.6,
after the major merger epoch.  This latter ratio also holds for the 
entire subhalo population, as we see from the black solid curve at \rvir. 
Hence, there are more, by a factor of $\sim 2$, subhalos by number in the
innermost region of the BDM prime halo, but not by mass. Their DM mass is
comparable to that locked in the PDM subhalos. 

This excess of the subhalos within the inner 30~kpc of the prime BDM
halo points to a shorter capture time of these objects in the central region
than in the PDM model (as shown in section~4). This in turn leads to shorter 
lifetimes of the BDM subhalos.  On the other hand,
the total DM masses of PDM and BDM populations in the central region
are comparable. Partially this is because
a fraction of DM, $\sim 17\%$, has been converted into baryons
in the initial conditions. As a result, a larger fraction of the  BDM 
subhalos falls below the threshold of $N_{\rm DM}=100$, and, 
therefore, does not contribute to the subhalo mass in our estimates.
On top of this, because the BDM subhalos concentrate deeper in the
potential well of the prime halo, they are tidally stripped at a
faster rate and hence have lesser masses on the average.

\subsection{Subhalo Population Spatial Distribution}

\begin{figure*}
\begin{center}
\includegraphics[angle=0,scale=0.4]{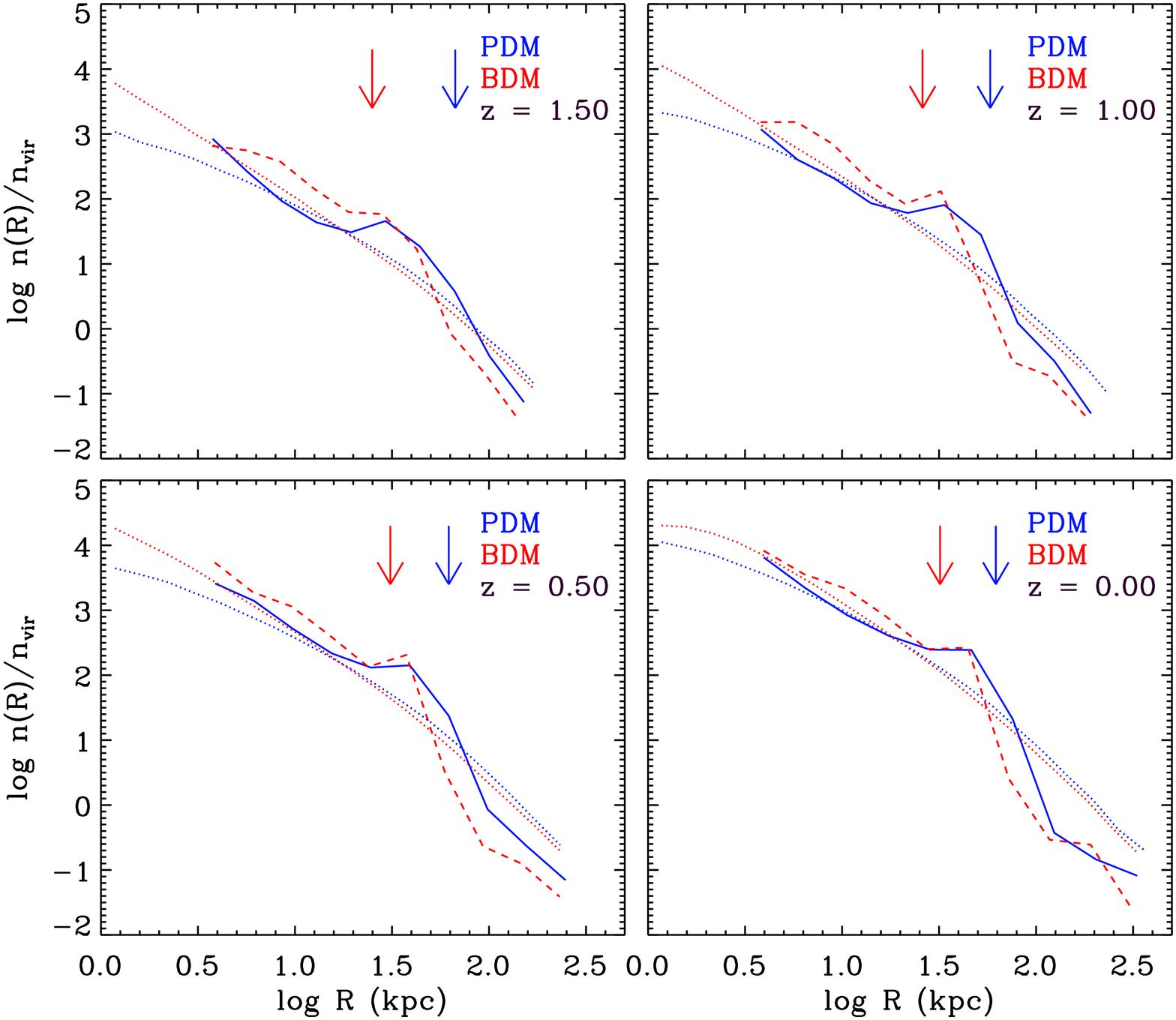}
\end{center}
\caption{Evolution of the subhalo population DM radial number density 
distributions within the prime halos in the PDM (blue solid) and BDM 
(red dashed) models, for subhalos with $N\ge 100$ or $M_{\rm sbh} \gtorder
2.7\times 10^8~{\rm M_\odot}$. The number density, $n(R)$, is normalized
by the average value of subhalo population DM density within \rvir. 
The prime halos DM density profiles are shown for comparison (same
colors, dashed lines).
Snapshots are shown for four different $z$.  The vertical arrows correspond
to the instantaneous values of \rvmax{} (PDM and BDM) in these models
with corresponding colors. All curves have been terminated at the
instantaneous values of \rvir{} for the prime halos.  }
\end{figure*}
\begin{figure*}
\begin{center}
\includegraphics[angle=0,scale=0.4]{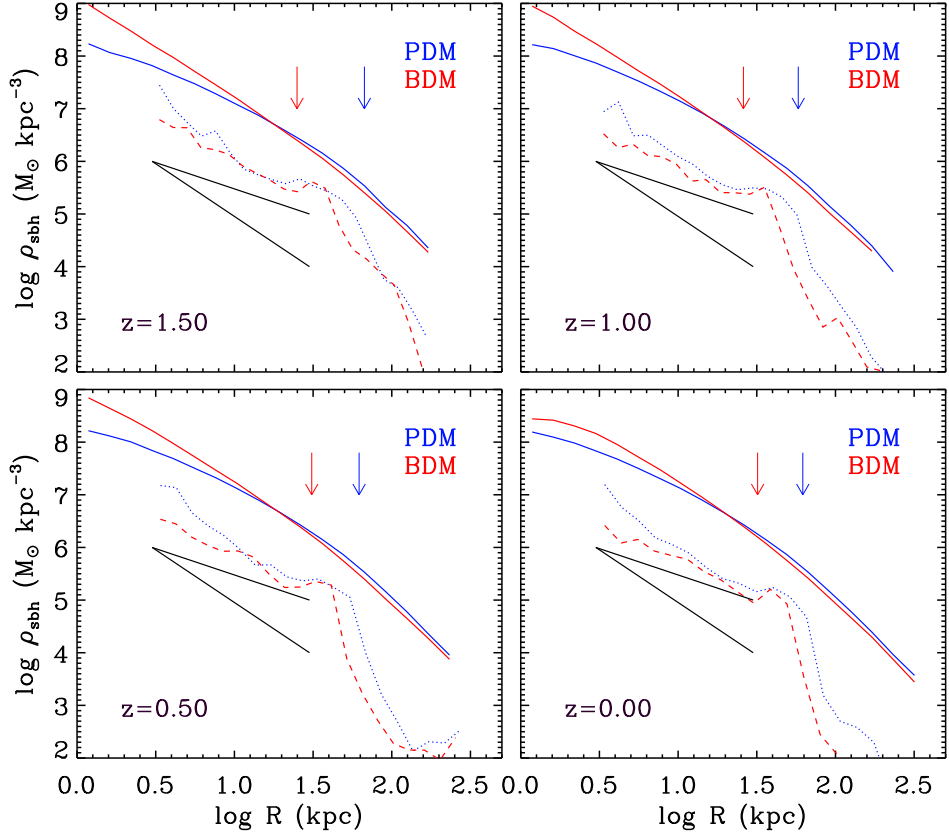}
\end{center}
\caption{Evolution of the subhalo population DM radial mass 
distributions 
within the prime halos in the PDM (blue dashed) and BDM (red dashed)
models, for subhalos with $N\ge 100$ or $M_{\rm sbh} \gtorder
2.7\times 10^8~{\rm M_\odot}$. Similar but normalized number density
distributions are shown in Fig.~4. Snapshots are shown for four different
$z$. The prime halos DM mass distribution is given by the solid blue
(PDM) and red (BDM) lines. Black solid lines show the -1 slope (upper)
and -2 slope (lower) for a comparison. The vertical arrows correspond
to the instantaneous values of \rvmax{} (PDM and BDM) in these models
with corresponding colors. All curves have been terminated at the
instantaneous values of \rvir{} for the prime halos.  }
\end{figure*}

\begin{figure}
\begin{center}
\includegraphics[angle=0,scale=0.41]{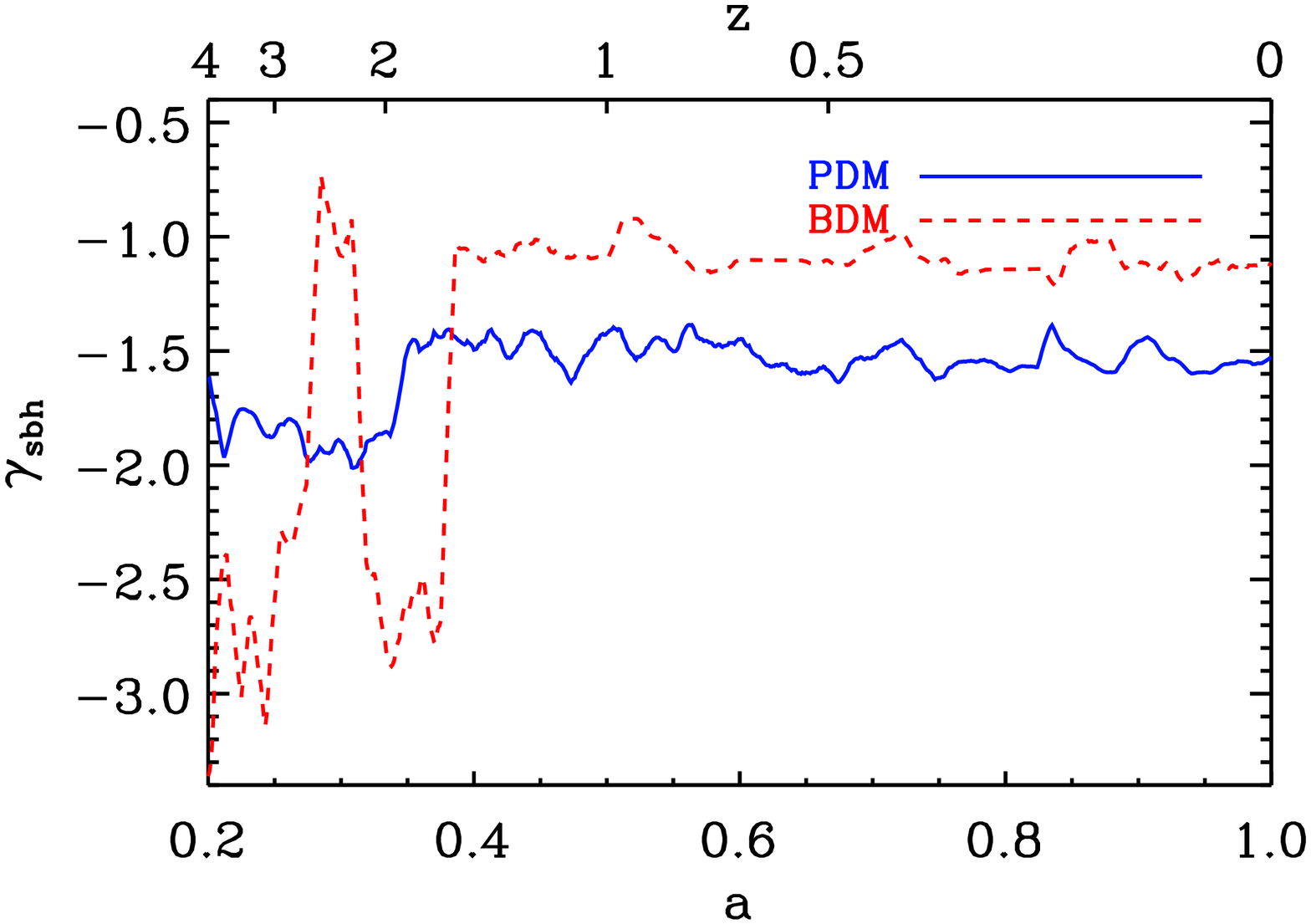}
\end{center}
\caption{Evolution of subhalo population DM density profiles power
index $\gamma_{\rm sbh}$, within \rvmax{} shown in Fig.~5.  The
density profile is approximated by a power law, $\rho_{\rm sbh}\sim
r^{\gamma_{\rm sbh}}$ for PDM (solid blue) and BDM (dashed red)
models.  The time averaging involves 20 frames.  }
\end{figure}

As a next step, we have computed the number density of subhalos, 
$n(R)/n_{\rm vir}$, normalized by their average density within \rvir,
in the PDM and BDM models (Fig.~4). As before, we only account for 
subhalos with a mass above the threshold mass of 
$M_{\rm sbh}\gtorder 2.7\times 10^8~{\rm M_\odot}$ (i.e., 
$N_{\rm DM}\ge 100$). The prominent feature in $n(R)$ is a 
bump associated with some steepening of the slope at \rvmax.
As a next step, we analyze the DM mass distribution due to the 
subhalo population, $\rho_{\rm sbh}$, within the prime halos of both
models. We find that the subhalo population
distribution bears some similarity to that of the total DM mass
distribution in the prime halos but also shows some differences. Within
\rvmax, the DM density decrease appears linear in the log~$\rho_{\rm
sbh}$---log~$R$ plot for both subhalo populations. We observe a sharp
steepening of the density around \rvmax{}. This characteristic radius
is always larger in the PDM model by a factor of 2 (see Fig.~4 and
Paper~I). The subhalo population densities are typically steeper than
-2 slope beyong this radius. We can trace the subhalos by almost a decade
in radius inside \rvmax, to the inner few kpc --- the resolution limit
of the subhalo detection. Within \rvmax{} the slopes lie
between -1 and -2, as seen in the snapshotes in Fig.~4 at various $z$,
with the PDM slope being somewhat steeper on the average
(more about this below).  As confirmed by the lower frame of Fig.~3,
the rate of the subhalo influx into the central region is uneven which
leads to an alternate dominance by the PDM or BDM subhalos at various
times.  Accordingly, the same behavior is found when we divide the
subhalo population into spherical shells of equal width.  

\begin{figure*}
\begin{center}
\includegraphics[angle=0,scale=0.403]{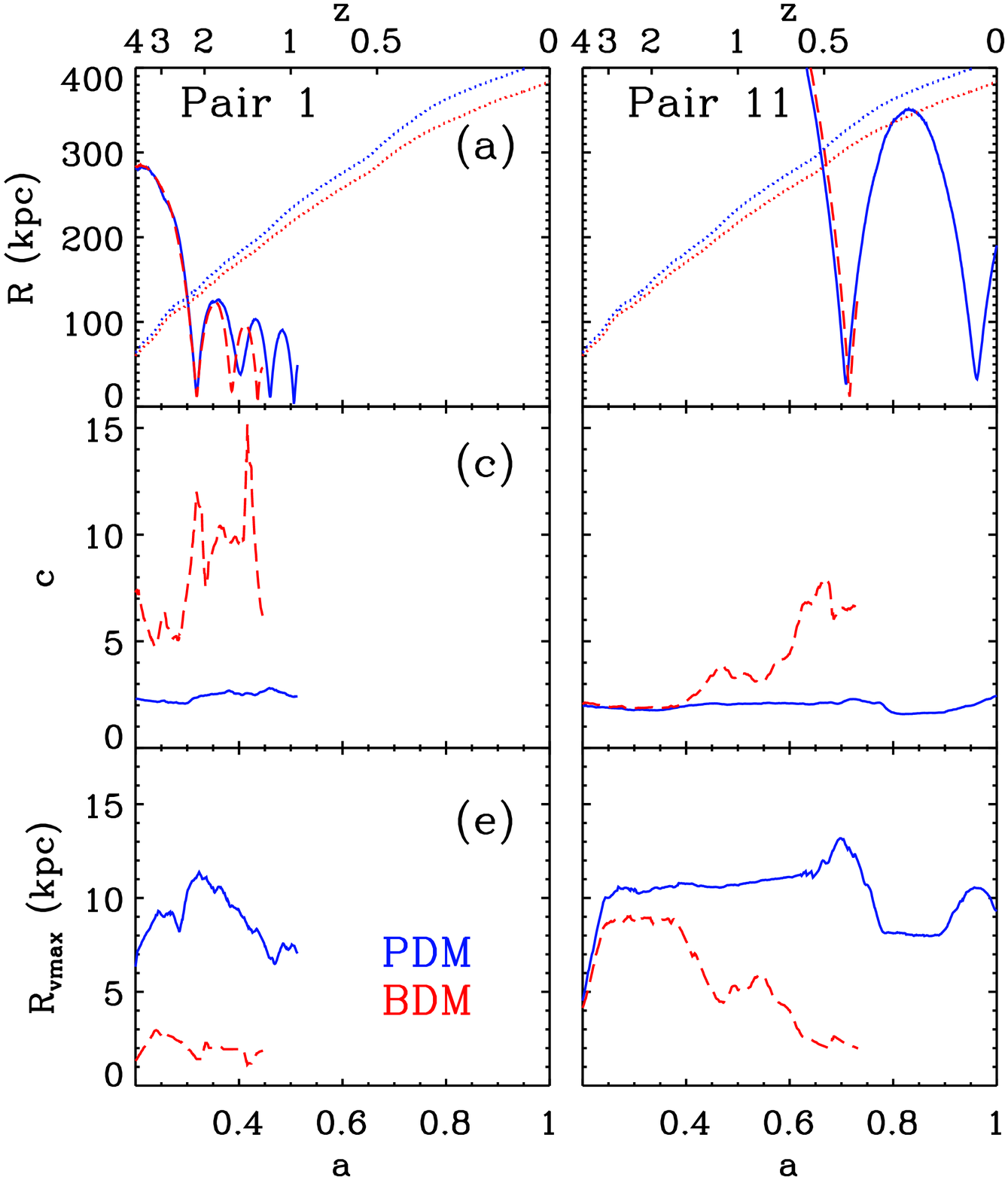}
\includegraphics[angle=0,scale=0.403]{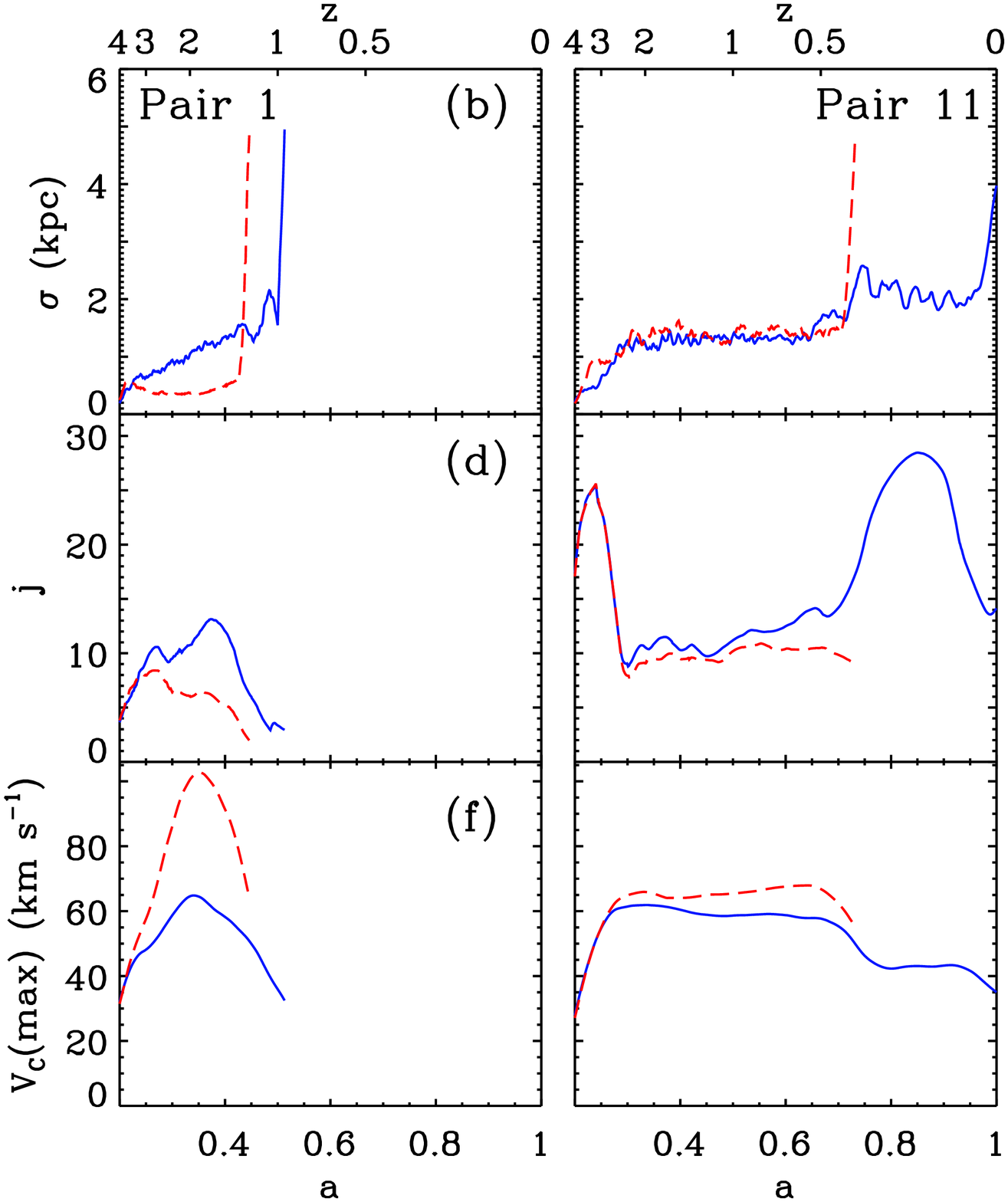}    
\end{center}
\caption{Examples of evolution of corresponding subhalo pairs in PDM
and BDM models. Shown are $(a)$ the radial trajectories given as the
position of a subhalo with respect to the CoM of the prime halo, $(b)$ 
relative distances of the 20 densest particles, $(c)$ concentration 
parameters, $(d)$ specidic angular momenta with respect to the prime 
halo CoM, $(e)$ \rvmax, and $(f)$ circular velocities at \rvmax. 
\underline{\it Left frames:} subhalo pair \#1,
\underline{\it Right frames:} subhalo pair \#11.}
\end{figure*}

We have fitted a time-averaged power law to the density profile of
subhalo population within \rvmax, $\rho_{\rm sbh}\sim r^{\gamma_{\rm sbh}}$ 
shown 
in Fig.~5 (see Fig.~6). The PDM and BDM distributions differ,
the former being somewhat steeper, with $\gamma_{\rm sbh} \sim -1.5$ vs $-1$ for
the BDM subhalos, confirming Fig.~5. We note also that the profiles
are more stable than those for the SubHMF, after the epoch of
major mergers.  The slopes are shallower than those of the total
DM distribution in the prime halos within \rvmax{} and steeper outside
these radii. This behavior can be dominated by dissolution of the subhalos
in the central region.

We have calculated the DM mass fraction of subhalo populations locked
within \rvir{} of the prime halos. After $z\sim 3$, both the PDM and
BDM models behave similarly. This confirms our previous estimates of
the DM associated with subhalos and with their tidally disrupted
remanants (i.e., streamers, still visible in the phase space) within
the prime halos (Fig.~11 of Paper~I). In Paper~I we have estimated
that the total fraction of bound (to subhalos) and unbound but not yet
mixed material at $z=0$ to be $\sim 8\%-9\%$. Here we find that about
half of this DM is bound and the rest contributes to the streamers,
i.e., to the not yet mixed material.

We also comment on the axial ratios of the subhalo populations. As shown
in Fig.~1, these ratios are much closer to unity in the presence of
baryons. This agrees well with our results for the DM distribution
in the prime halos in paper~I.

\section{Results: Subhalo Populations in PDM and BDM Models}
 
Using the compiled catalog of subhalos in PDM and BDM models allows us
to make a direct comparison between corresponding pairs in order to
understand the effect of baryon presence on the subhalo population and
its dynamics. Specifically, we have compared the corresponding
subhalos trajectories, masses within their \rt{} and \rvmax, their
concentration parameters, masses and mass ratios of all components to
their total masses, specific angular momenta, and determined their
destruction times. The redshifts of the prime halo \rvir{} crossing
have been choosen to lie between $z\sim 4-0.5$. Two examples from this
catalog are shown in Fig.~7 and provide a detailed quantitative
analysis of these subhalo pairs. The subhalo population average
properties are discussed later on.

Analyzing the corresponding PDM and BDM pairs, we find that the
redshift of a subhalo disruption, $z_{\rm des}$, is almost always
higher for BDM than for the associated PDM ones. This seems to be
counter-intuitive because one can argue that the BDM subhalos are more
concentrated due to the baryon dissipation. This argument, however,
neglects the accompanying changes to the BDM prime halo, which becomes
substantially more concentrated, as shown in Fig.~4 of Paper~1.  The
orbits of PDM and BDM subhalo pairs appear nearly identical up to
their first pericenter.  After this, the trajectories separate
(Fig.~7a). The radial extent of the PDM trajectories remains larger,
and the next apocenter of the PDM subhalo lies at larger $R$ compared
to the BDM. We also find that after the first pericenter is reached,
the specific angular momentum of a PDM subhalo's
particles is always larger than that of the corresponding BDM, with respect
to the CoM of the prime halo (Fig.~7d). In other words, the BDM subhalos
settle on more radial orbits after the first pericenter passage. 
The tidal disruption of
subhalos normally occurs after the pericenter, as can be seen from the
suddent increase in $\sigma(z)$ (Fig.~7b). The larger time spacing
between the pericenters of the PDM subhalos is probably the main
reason why the PDM population is characterized by a longer lifetime
compared to their BDM counterparts. We return to this issue later on
when we discuss the average properties of subhalo populations.

\begin{figure*}
\begin{center}
 \includegraphics[angle=0,scale=0.91]{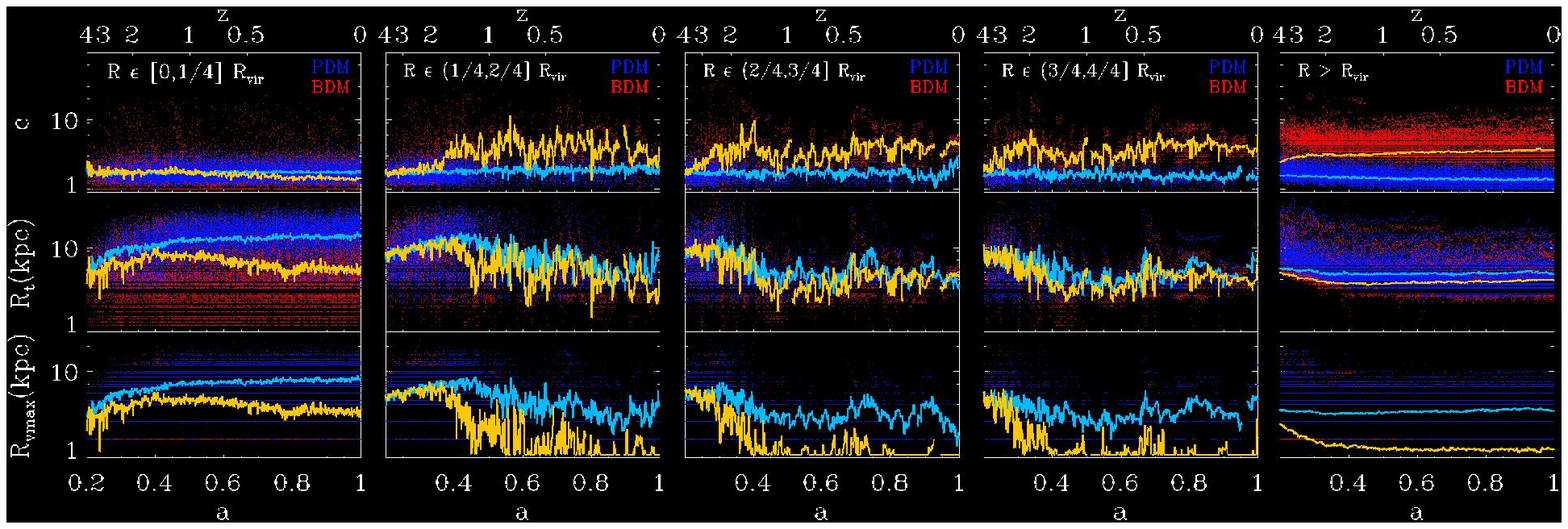}
\end{center}
\caption{Evolution of the concentration parameter $c$, \rt{}
and \rvmax{} in PDM (blue) and BDM (red) subhalos within five radial
zones: $r=0-0.25$\rvir, $r=0.25-0.5$\rvir, $r=0.5-0.75$\rvir,
$r=0.75-1$\rvir{} and $r>$\rvir{} (from left to right), where \rvir{}
is that of the prime halo. The mean values for each population are
shown as a yellow line (for population of red dots) and a light blue 
line (for population of blue dots). }
\end{figure*}

\begin{figure*}
\begin{center}
\includegraphics[angle=0,scale=0.32]{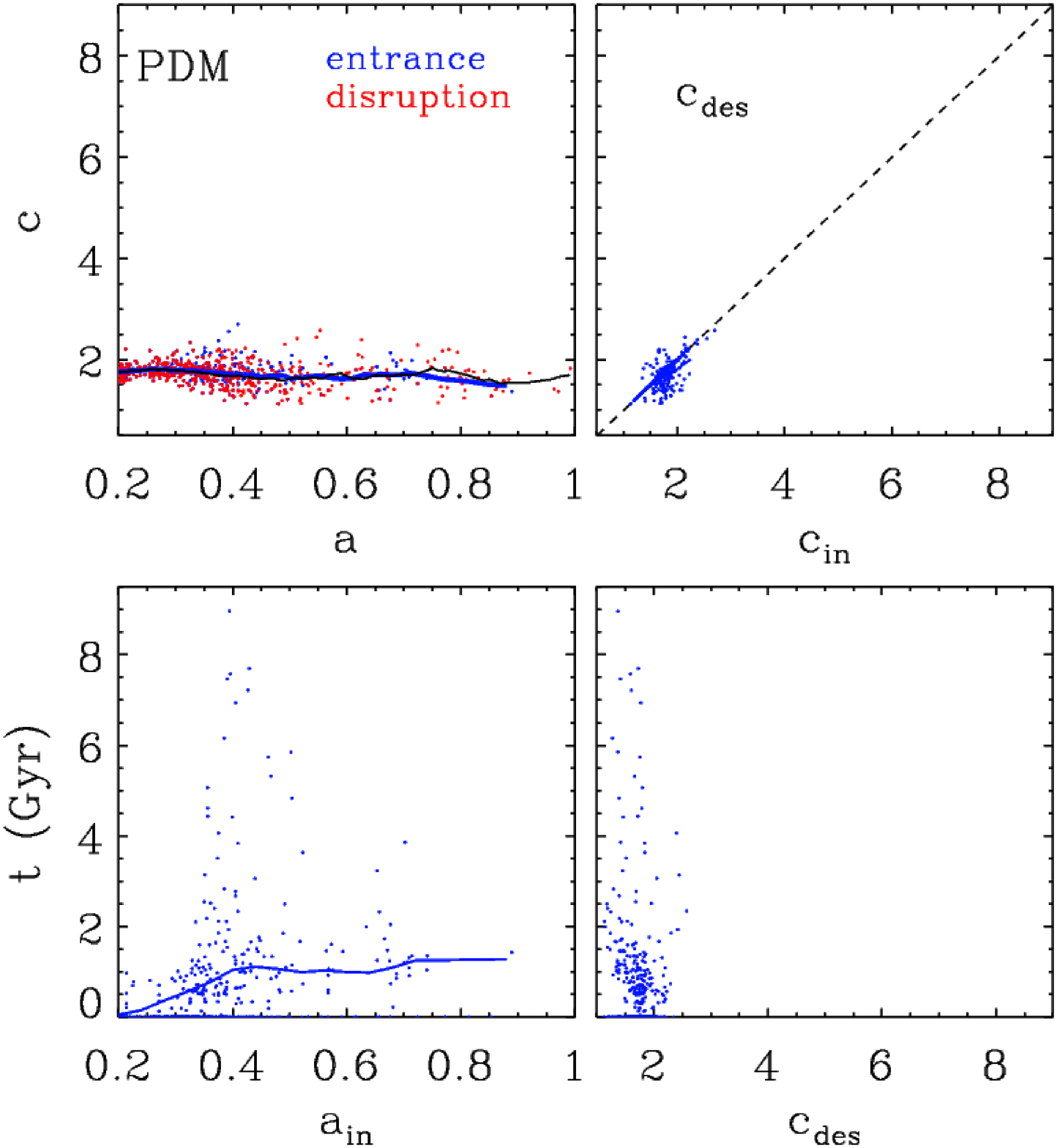}
\includegraphics[angle=0,scale=0.32]{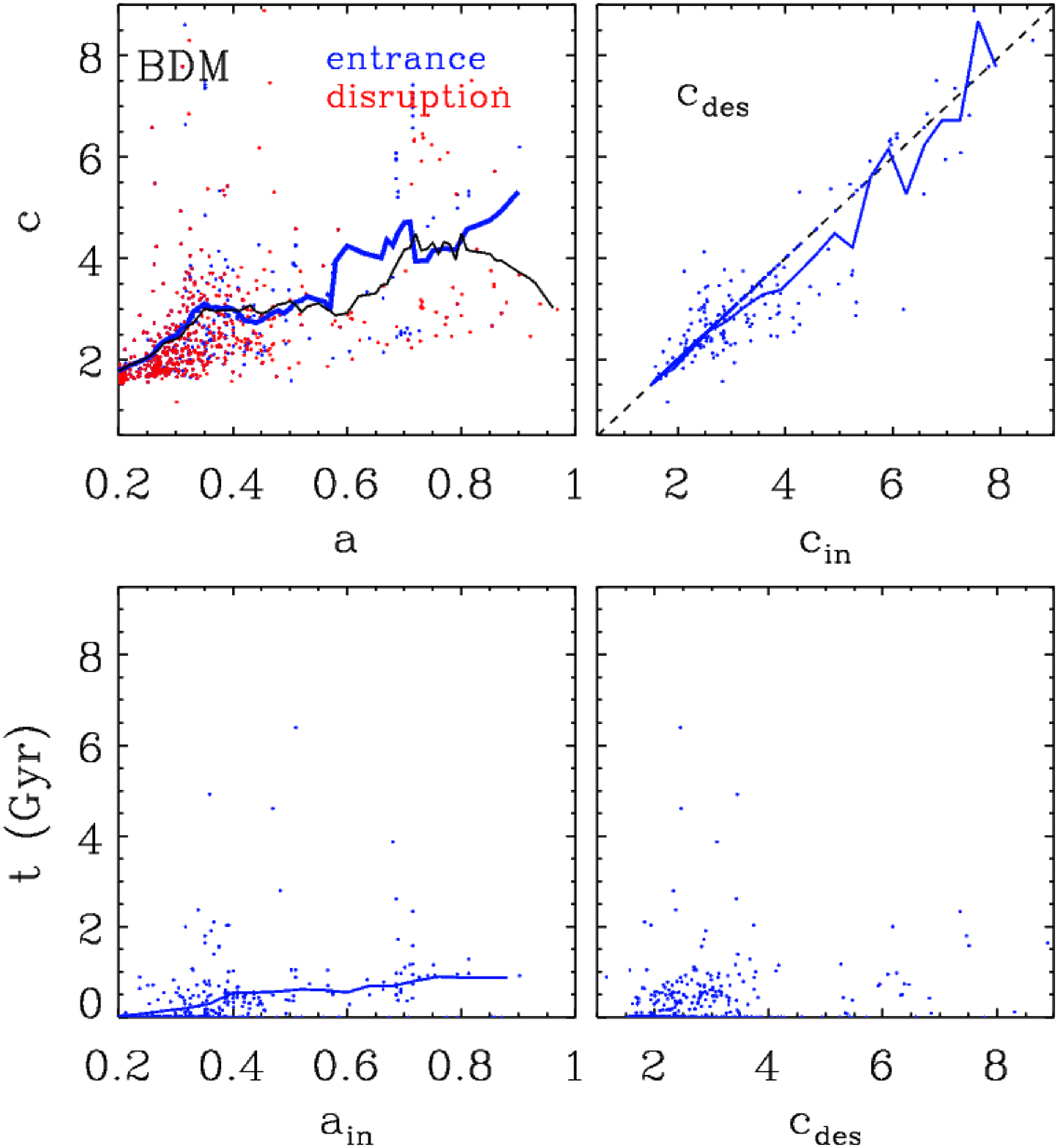}
\end{center}
\caption{Evolution of the subhalos properties in PDM (left frames) and
BDM (right frames). {\it Upper left:} $c$ as a function of $a$; the
average values of $c$ at the time of entering the prime halo (blue
solid) and being tidally disrupted (red solid) are shown as well. {\it
Upper right:} concentration parameter at the time of the tidal
disruption $c_{\rm des}$ vs concentration parameters at time of entrance, 
$c_{\rm in}$; the
diagonal dashed line is shown for comparison.  {\it Lower left:}
lifetime of subhalos as a function of $z$ and $a$; the average value
are shown a solid blue line. {\it Lower right:} lifetime of subhalos
as a function of $c_{\rm des}$. Only subhalos that are destroyed
before $a=1$ are shown in these diagrams.  }
\end{figure*}

The next important difference between the corresponding PDM and BDM
subhalos is the evolution of the concentration parameter. Already at
higher $z$, close to the turnover radii, when the subhalos travel
outside \rvir{} of the prime halo, $c$ grows faster in BDM
subhalos. This growth seems spectacular when compared to a nearly
constant $c$ of the PDM counterparts (Fig.~7c). It terminates
typically when a subhalo enters \rvir{} of the prime halo, or after
the first pericenter of its orbit.  However, the cutoff in the growth
of $c$ can come as early as the turnover radius of the
prime. Frequently, the ``normal" evolution of $c$ is modified by
interactions with other subhalos.

Before we turn to the average properties of the subhalo populations,
we note that the reason for consistently higher $c$ in BDM subhalos is
that their \rvmax{} decreases early and substantially due to the
accretion of baryons which concentrate deep within this radius
(Fig.~7e), as can be confirmed by comparing the radial profiles of
their circular velocities and by the maximal values of circular
velocities (Fig.~7f). Hence, baryons clearly impact the internal
structure of the BDM subhalos. We note that \rvmax{} in the PDM subhalos
declines late in the evolution as well, typically in the last
stage leading to a tidal disruption (Fig.~7e). This effect was first
noticed by Klypin et al. (1999). However, $c$ stays nearly constant
in these objects, as the decline in \rvmax{} is concurrent with the
decline in \rt.

In order to follow the overall properties of subhalo populations in
both models, we separate their evolution into 4 spatial zones within
and one zone outside the prime halos, namely, $0-0.25$\rvir,
$0.25-0.5$\rvir, $0.5-0.75$\rvir, $0.75-1.0$\rvir{} and $>$\rvir{},
where the virial radius is that of the prime halos (Fig.~8).  In the
innermost region, the average $c$ remains nearly constant in time for
the PDM and slowly decreases for the BDM subhalos, below the PDM
curve. For the next three regions, the average $c$ is somewhat more
noisy for the PDM and much more noisy for the BDM ones. The overall
trend is of a nearly constant $c$ for PDM and a slight increase and a
subsequent decrease for the BDM average values. However, the BDM curve
always stays above the PDM one in the next three zones. This trend
continues outside \rvir, although the noise has disappeared completely
and the separation between the average curves is larger. Strikingly,
the dispersion of $c$ values around the average is much larger for the
BDM model in all zones, especially in the innermost one.

As has been noted already in Fig.~7 for specific examples and
confirmed in Fig.~8 for the whole subhalo population, the divergent
behavior of the average $c$ comes from a decrease in the average values
of \rt{} and especially of \rvmax{} in the BDM subhalos. The maxima of
the circular velocity curve move to much smaller radii in the BDM
subhalos due to the captured baryons contribution. This is most
dramatically demonstrated for the substructure outside the prime halo
--- here \rvmax{} drops to a few kpc and stays nearly constant after
the major merger epoch, while that of the PDM population is barely
affected.

To further compare the evolution of $c$ for the PDM and BDM subhalos,
we plot its values for individual subhalos at the time they enter
\rvir{} of the prime halo and at the time they are tidally disrupted,
$c_{\rm in}$ and $c_{\rm des}$ respectively, vs time (Fig.~9, left
upper). The prevailing trends are obvious for both populations: no
evolution for the PDM and a gradual increase by a factor of $\sim 2-3$
in BDM. Note that occasionally, the entering $c$ exceeds that of the
disruption, but overall they evolve in tandem.

Because little evolution in $c$ is observed in PDM, the $c$ values lie
in the corner of the available parameter space, $c_{\rm des}-c_{\rm
in}$, in PDM, but are much less confined in the BDM case (Fig.~9,
upper right). Again we confirm that the average ratio of $c_{\rm
des}/c_{\rm in}$ is nearly constant for both populations.

Comparison of the subhalo lifetimes following their entrance into the
prime halos shows that the average time before the tidal disruption
increases in time during the major merger epoch, then nearly levels
off. The average lifetime thereafter is about 1.5~Gyr for the PDM and
$\sim 0.9$~Gyr for the BDM subhalo population. However, the dispersion
around this mean is substantial, especially during the major mergers
epoch. For PDM, most of the subhalos are destroyed within 4~Gyrs,
some survive much longer, $\sim 5-9$~Gyr (Fig.~9, lower left)). The
lifetime of the BDM subhalos is much more limited, and most are
destroyed within $\sim 1-2$~Gyr.

Finally, we find that, on the average, the PDM subhalos are destroyed
within a narrow range in $c_{\rm des}$, while the BDM ones exhibit a
much larger dispersion in these values (Fig.~9, lower right). We also
find a clear symmetry across the $c_{\rm in} - c_{\rm des}$ diagonal
in both models --- thus the subhalos do not increase their $c$ as they are
tidally disrupted. This is interesting because
Fig.~7c shows that $c$ in BDM increase with time after penetrating the
prime halo, although they fall sharply during the tidal disruption
process.

Comparison with the PDM model reveals an important difference ---
although the BDM subhalos are more centrally concentrated than their
PDM counterparts, they are disrupted at a higher rates. This applies of
course only to the substructure that penetrates deep inside the prime
halo, say the central $\sim 30$~kpc.

\section{Discussion and Conclusions}

The purpose of this work is to compare the properties and evolution of
subhalos without baryons (PDM models) and in their presence (BDM
models). PDM and BDM models have been evolved from identical initial
conditions designed by means of Constrained Realizations method. The
prime halos of both models have been compared in paper~I. We have
compiled a comprehensive catalog of PDM and BDM subhalos in order to
analyze the statistical properties of these populations. In addition,
we created a subsample of corresponding PDM and BDM subhalo pairs to
focus on the individual evolution. Lastly, we have performed some basic
analysis of subhalo population orbits in PDM and BDM.

Our main results are as following. We compile and compare the subhalo
population mass functions (SubHMFs) in the range of $z\sim 3-0$.  The
SubHMFs are consistent with a single power law in the mass range of
$\sim 2\times 10^8~{\rm M_\odot} - 2\times 10^{11}~{\rm M_\odot}$ for
both models. However, we note a systematic offset of the BDM
population with respect to the PDM one. Namely, $<\alpha>\sim -0.86$
(PDM) and --0.98 (BDM), respectively. Some variablity in $\alpha(z)$
can come from an intrinsic scatter of $\pm 15\%$. In the simulations
it is associated with waves of subhalos falling into the prime halos
along the large-scale filaments. Such waves appear to be much more
pronounced in the presence of baryons.

Next, we have analyzed the spatial DM number and mass densities of 
subhalo populations. We find that the latter one can be roughly 
described as a 
power law with a steepening around \rvmax{} --- the radius of
maximal circular velocity and hence largest binding energy. Inside
\rvmax{} the slope, $\gamma_{\rm sbh}$, is much more shallow than
outside this radius. A smoothed out $\gamma_{\rm sbh}$ is about --1.5
in the PDM and --1 in the BDM models, within \rvmax. The slope is
closer to --3 outside this region. The shallower slope within \rvmax{}
is related to the elevated rates of subhalo ablation and tidal
disruption processes. Overall, the subhalo number density profiles 
follow the distributions of their respective main halos.

Third, we find that the BDM population of subhalos is depleted at a
higher rate compared to the PDM population. This result appears to be
counter-intuitive because one expects the accreted baryons to increase
the mass concentration of BDM subhalos and make them more resilient to
tidal disruption. However, we find that the trend leading
to the increased mass concentration of the prime halo in the BDM model
dominates and ultimately contributes to the dissolution of subhalos.
Specifically, trajectories of PDM-BDM subhalo pairs appear to {\it diverge}
after the first pericenter. The subsequent BDM subhalos trajectories
become confined to the innermost regions of the prime halo where they
experience increased dynamical friction and spiral in towards their
ultimate tidal disruption. The  lower slope for the inner density 
profile of the BDM subhalos compared to the PDM stems from their 
elevated rate of the tidal disruption. The small number of central 
subhalos leads to significant fluctuations in the slope. Yet, on the 
average, the BDM population is depleted faster in this region as 
shown by Figures 6 and 9, resulting in a lower slope.

We now compare our results with those found in the literature, starting
with the SubHMFs. Most of the modeling of SubHMFs on the galactic scales 
has been focused on the pure DM models. Comparison with the BDM models
is nearly always limited to galaxy clusters. In the galaxy and cluster mass 
range, the slope
$\alpha$ was found to be consistent with -1, for the PDM models (Moore 
et al. 1999; Ghigna et al. 2000; Springel et al 2001; Stoehr et al. 2003; 
De Lucia et al. 2004; Gao et al. 2004; van den Bosch et al. 2005; 
and Diemand et al. 2007, 2008). More recent multi-resolution
Aquarius simulation (Springel et al. 2008) estimate the range for 
$\alpha$ to lie between  -0.87 and -0.93, close to our value of -0.86.

The later evolution of the PDM SubHMFs has also been noticed in the Via
Lactea simulation. For via Lactea and additional simulations, Madau 
et al. (2008) showed that the slope of the mass function at two
different redshifts ($z = 0.5, 0$ --- the only ones published) has 
slightly changed, steepening  from $-0.92$ to $-0.97$. Such a ``last 
moment" steepening is compatible with our Fig.~2, as noted in section~3.1.
We find that this possible steepening in PDM and BDM subhalo populations
can be related to the small number statistics toward the end of the
simulation --- i.e., we possibly observe a large amplitude variation
around the mean for the slope $\alpha$. We, therefore, leave this issue
open.

There is much less data on the SubHMF in the presence of baryons, and it
deals with the mass range of clusters of galaxies. Comparing 
simulations (with and without baryions) drawn from the same initial 
conditions, Weinberg et al. (2008) found the two subhalo mass functions 
to be very similar. They conclude that the
dissipative baryonic component has only a ``small'' impact on this
global measure of the subhalo population. On the other hand, Dolag et al. 
(2009) find that the addition of baryons somewhat modifies the SubHMF for
galaxy clusters, with $\alpha_{\rm PDM}$ to be somewhat steeper than 
$\alpha_{\rm BDM}$. While their values of $\alpha$ are similar to ours, 
the order appears inverted. Finally, for $z=0$, Libeskind et al. (2010) 
point out that the BDM substructure is more radially concentrated 
than in the PDM.

Next, we turn to the DM density profiles of the subhalo population.
We find Maccio et al. (2006) as the only work on this issue within the galaxy
mass range, although they stopped the gas cooling at $z=1.5$
(see also Sales et al. 2007, although they only analyze the most
massive subhalos within their simulations). This work 
provides only the subhalo number density profile (their Fig.~6), which is 
consistent with ours. Similar results have been obtained by Nagai \&
Kravtsov (2005) and by Weinberg  et al. (2008) --- all deal with the
subhalo population number density profiles, albeit in the galaxy clusters
mass range. Furthermore, we find that there is little spatial bias (less 
than $10\%$) between the subhalo population DM mass density and the DM 
distribution of the main halos (e.g., Sales et al. 2007).

Extending the subhalo population number density to the DM {\it mass} density,
we find that a new feature --- steepening of the mass density profile, 
appears to be associated with the position of \rvmax{} in each model. 
Around \rvmax, the slopes change and steepen from $\gamma_{\rm sbh}\sim
[-1]-[-1.5]$ to below -2. If verified in other models, this is an
interesting effect related to increased ablation and dissolution of subhalos
in the cnetral region of the prime halos. It has implications on dynamics
of subhalos and reflects their trapping within \rvmax.
 
Turning to the properties of individual subhalos in PDM and BDM models, 
specifically to the evolution of the concentration parameter, $c$, we
find that except in the innermost 30~kpc the average $c$ of the BDM
population is higher than that for the PDM one (Fig.~8). Hence, it seems 
surprising that both PDM and BDM subhalos have the same $c$ when they
enter the prime halo and when they are tidally disrupted, i.e.,
$c_{\rm des}\approx c_{\rm in}$, as shown by Fig.~9. Resolution of this
puzzle is provided by Fig.~7c. Indeed, $c$ in BDM stays well above that
for PDM and increases with time. However, in the process of a tidal
disruption, it drops sharply, and in many cases to near the original
value, thus assuring that the population average has 
$c_{\rm des}\approx c_{\rm in}$. This coincidence is related to our
choice of the ``moment of disruption" --- taking $\sigma=5$~kpc for the
dispersion in the positions of 20 densest particles in a subhalo. 
However, this definition has been tested, including visual test as 
well as a test in the phase space, and found highly reliable. Any
other choice of the critical value for $\sigma$ would be more {\it ad 
hoc}. 
 
Our result for the average $c$ for the subhalos population in both models
agree well with the conclusion by Dolag et al. (2009) that BDM population
is on the average more concentrated than the PDM, albeit they show only a
single snapshot in redshift and in the cluster regime. They also use the
standard definition of $c$ by fitting the NFW profile to subhalos.
However, we find that the subhalos are tidally truncated and avoid
using the NFW model fit. The tidal effects are especially visible in
the central region of 30~kpc where we find that the PDM are on the average
less concentrated than the BDM ones. This of course must be taken
together with the caveat that the PDM subhalos spend less time
inside this region, as discussed above (see also Fig.~7a).

This brings us to an interesting point, namely, which subhalo population 
is tidally disrupted faster. While it is obvious that baryons provide an
additional ``glue" to the DM subhalos, they also affect the innermost
region of the prime halo. This is obvious from our analysis of DM
distribution in the prime halo of the BDM model (Paper~I), where \rvmax{} 
is essentially halved compared to the PDM prime. Even more prominent 
changes are affecting the cusp region (Romano-Diaz et al. 2008a).
Clearly, the resolution of the question of which population is destroyed
at a higher rate depends on the competition between two effects: increase
in the concentration of the subhalos versus increase in the concentration
of the prime halos. In our simulations, the BDM population appears to be
destroyed at  a higher rate in the central region. However, if the
energy and momentum feedback from stellar population and from the central
supermassive black hole are able to decrease and maintain a lower
fraction of baryons in the prime halo, the effect on the DM density
profile will be sharply decreased and the BDM subhalo population will
be much more resilient to destruction than the PDM one. It is possible
that nature benefits from both solutions. In addition, it is 
plausible that the galaxy cluster environment prefers the survival of 
the BDM subhalos, (as in Weinberg et al. 2008; Dolag et al 2009), while 
on the smaller mass scale, the galaxy environment lead to the opposite. 

With respect to the total fraction of the halo mass invested within
substructure we find that $\sim 5\%$ is locked within subhalos. This
result is consistent with previous published analysis (e.g., Ghigna et 
al. 2000; Springel et al. 2001; Stoehr et al. 2003) which provide
a range between 5\% and 20\%. However, there is no agreement on
this matter, as Moore et al. (2001) argues that the true fraction might
approach unity if the subhalos could be identified down to very small
masses. Our results are also consistent with the Via Lactea simulations
where the subhalo mass fraction is $\sim 5.3\%$ within \rvir, but 
appear somewhat lower compared to Aquarius simulation, which quotes 
$\sim 11.2\%$.

To summarize, we have analyzed the properties and evolution of subhalo
population in the mass range of $10^8~{\rm M_\odot}-10^{11}~{\rm M_\odot}$
in pure DM and DM$+$baryons models evolved from identical
initial conditions. Our main results are that the subhalo mass functions
can be fitted with a single power law which is compatible with no
redshift evolution between $z\sim 3-0$, but which is somewhat steeper
in the presence of baryons. We have also computed the number and mass
density profiles for the DM subhalo component in both models. We find that
the DM mass density is shallower inside the radius of the maximal
circular velocity in the prime halos and steepens outside. Finally,
we compared the disruption rates for the subhalo populations and
find that, in the presence of baryons, the subhalos are destroyed
at a higher rate within the central 30~kpc of the prime halo --- a
direct result from the increased central mass concentration. However,
with a larger feedback from stellar populations and the central
supermassive black holes, the effect can be reversed.

\acknowledgments 
We are grateful to our colleagues, too numerous to list here, for discussions 
on various topics addressed here. This research has been partially supported 
by NASA/LTSA/ATP/KSGC, STScI and the NSF grants to I.S. STScI is operated 
by the AURA, Inc., under NASA contract NAS 5-26555. I.S. is grateful to the 
JILA Fellows for support. Y.H. has been partially supported by a grant 
from the ISF (13/08). C.H. has been partially supported by a grant from the NSF.
  


\end{document}